# Angular surface solitons in sectorial hexagonal arrays

A. Szameit[1], Y. V. Kartashov[2], V. A. Vysloukh[2], M. Heinrich[1], F. Dreisow[1],T. Pertsch[1], S. Nolte[1], A. Tünnermann[1,3], F. Lederer[4], and L. Torner[2]

[1]*Institute of Applied Physics, Friedrich-Schiller-University Jena, Max-Wien-Platz 1, 07743 Jena, Germany*

[2]*ICFO-Institut de Ciencies Fotoniques, and Universitat Politecnica de Catalunya, Mediterranean Technology Park, 08860 Castelldefels (Barcelona), Spain*

[3]*Fraunhofer Institute for Applied Optics and Precision Engineering, Albert-Einstein-Strasse 7, 07745 Jena, Germany*

[4]*Institute for Condensed Matter Theory and Optics, Friedrich-Schiller-University Jena, Max-Wien-Platz 1, 07743 Jena, Germany*

We report on the experimental observation of corner surface solitons localized at the edges joining planar interfaces of hexagonal waveguide array with uniform nonlinear medium. The face angle between these interfaces has a strong impact on the threshold of soliton excitation as well as on the light energy drift and diffraction spreading.

*OCIS codes: 190.0190, 190.6135*

Over the last several years there has been growing interest to linear and nonlinear optical surface waves. After the prediction of discrete surface solitons at the interface of uniform material and waveguide array with focusing nonlinearity [1] and at the interface between dissimilar arrays [2] and their subsequent observations [3-5], a burst of new findings revolutionized the perception of nonlinear surface waves. For example, solitons were also found at the edge of defocusing lattices [6-8]. Special attention was given to two-dimensional (2D) geometries [2,9-11]. 2D surface solitons were observed in optically induced lattices [12] and in laser-written waveguide arrays [13,14]. However, up to now the impact of the geometry of the interface on the properties of 2D surface solitons was not addressed properly. To bridge this gap, we report in this Letter the salient features of 2D solitons localized at the edges joining planar interfaces of a hexagonal waveguide array with a uniform nonlinear medium and show that the face angle is the key parameter defining the power threshold for soliton excitation and the direction of the light energy drift and diffraction spreading.

Such corner solitons offer additional opportunities for evanescent optical fields control at the convex and concave interfaces.

For the theoretical description of surface wave formation we employ the nonlinear Schrödinger equation for the dimensionless field amplitude $q$ under the assumption of cw illumination:

$$i\frac{\partial q}{\partial \xi} = -\frac{1}{2}\left(\frac{\partial^2 q}{\partial \eta^2} + \frac{\partial^2 q}{\partial \zeta^2}\right) - q|q|^2 - pR(\eta,\zeta)q. \tag{1}$$

Here $\eta,\zeta$ are the transverse coordinates and $\xi$ is the longitudinal one; the parameter $p$ characterizes the refractive index modulation depth; the function $R(\eta,\zeta)$ describes refractive index profile in the array and can be represented as a superposition of Gaussian functions $\exp[-(\eta/w_\eta)^2 - (\zeta/w_\zeta)^2]$. The individual waveguides with spacing $d$ are arranged in a hexagonal structure where only those waveguides that are located within the face angle $\alpha$ between the planar interfaces are present. In the particular case of a hexagonal structure the symmetry dictates five different types of interfaces (or corners) with angles $\alpha = n\pi/3$, where $n$ varies from 1 to 5. In our simulations we set $w_\eta = 0.45$, $w_\zeta = 0.9$, and $d = 4$ in accordance with the transverse waveguide dimensions of $4.5 \times 9$ $\mu\mathrm{m}^2$ and the spacing of 40 $\mu\mathrm{m}$. The parameter $p = 2.8$ is equivalent to a refractive index change of $3.1 \times 10^{-4}$, while the nonlinearity is focusing and spatially uniform. In all cases we excite only the corner waveguides (see Fig. 1, showing a microscope image of hexagonal array as well as the excited waveguides).

The face angle $\alpha$ substantially affects the linear light propagation. The rate of discrete diffraction spreading for a light beam launched in a corner waveguide increases with growing $\alpha$, and reaches its maximal value in the uniform hexagonal array. This is caused by the larger number of adjacent waveguides accompanying an increase of $\alpha$, which results in a faster leakage of energy from the excited waveguide. Additionally, the light tends to drift from the exited corner waveguide into the array depth, but the rate of this process slows down with the growth of $\alpha$, in contrast to the rate of diffraction. The integral center of the linear beam always shifts along the bisecting plane of the face angle $\alpha$. Initially the transverse shift $S$ of the integral beam center along the bisecting line increases $\sim \xi^2$, but at larger distances one finds only a linear dependence $S \sim \xi$. For $\alpha = \pi/3$ and $2\pi/3$ the shift rate $dS/d\xi$ at larger distances is almost equal ($dS/d\xi = 0.223$), but it drops to $dS/d\xi = 0.168$ at $\alpha = \pi$ and to $dS/d\xi = 0.031$ at $\alpha = 5\pi/3$. Such a shift characterizes the diffraction anisotropy which appears due to the specific boundary conditions.

The face angle $\alpha$ also strongly affects the characteristic features of nonlinear excitations. Soliton solutions of Eq. (1) in the form $q = w(\eta,\zeta)\exp(ib\xi)$ can be characterized by the power $U = \int\int_{-\infty}^{\infty} w^2 d\eta d\zeta$. For any $\alpha$ the corner solitons exist for $b$ values above a cutoff $b_{co}$ (Fig. 2). The power of such states always exceeds threshold power $U_{th}$, since the dependence $U(b)$ is nonmonotonic. In the region $b_{co} < b \leq b_{cr}$ one has $dU/db \leq 0$ that implies instability of corresponding branch, while for $b > b_{cr}$ the solitons are stable, since $dU/db > 0$. For $b \to b_{co}$ solitons expand substantially across the waveguiding sector, acquire an asymmetric shape and weakly penetrate into the uniform medium [Figs. 2(a) and 2(c)], while far from cutoff the light localizes in the corner waveguide [Fig. 2(b)]. At $b \to \infty$ when soliton becomes very narrow and lattice role diminishes the power approaches that of Townes soliton, i.e. $U = 5.85$. Importantly, the threshold power of corner solitons is a monotonic function of the face angle $\alpha$. Increasing $\alpha$ results in increase of the threshold power, since for higher $\alpha$ the nonlinearity has to counterbalance the stronger discrete diffraction. Hence, the threshold power varies notably from $U_{th} = 0.622$ for $\alpha = \pi/3$ to $U_{th} = 0.741$ for $\alpha = \pi$ and $U_{th} = 0.809$ for $\alpha = 5\pi/3$. The difference in threshold powers is most pronounced for smaller angles.

This trend is clearly seen in experiments performed in a waveguide array written with femtosecond laser pulses (Coherent Mira/RegA; see [15] for details of fabrication). The topology of this array (Fig. 1) allowed us to include all five types of wedge-like interfaces (from $\alpha = \pi/3$ to $\alpha = 5\pi/3$ in a single structure, so that the results can be directly compared. The writing velocity was $2000\ \mu\text{m/s}$, which ensures a nonlinear coefficient similar to that in the bulk material [16]. The spacing $d = 40\ \mu\text{m}$ yields almost isotropic coupling between adjacent waveguides [13]. The length of the sample was $105\ \text{mm}$, and the transmission losses of a single waveguide were $< 0.4\ \text{dB/cm}$. For soliton excitation we used a Ti: Sapphire chirped pulse amplification laser system (Spitfire, Spectra-Physics) with a pulse duration of $150\ \text{fs}$ and a repetition rate of $1\ \text{kHz}$ at $800\ \text{nm}$. In Fig. 3 the observed light evolution in the wedge-like array with $\alpha = \pi/3$ is shown and compared to the simulations according to Eq. (1). Note that experimental images taken upon excitation of different waveguides were rotated in order to have face angle increasing counterclockwise for different arrays. Due to the small face angle between the interfaces, the light penetrates notably into the array depth and the displacement of the integral beam center is rather strong at small power levels [Fig. 3(a)]. With increasing power, the output intensity distribution contracts towards the corner waveguide [Fig. 3(b)], so that at sufficiently high powers the drift and diffraction spreading are suppressed and a discrete corner soliton forms [Fig. 3(c)]. Since in this specific case ($\alpha = \pi/3$) the excited waveguides has only three neighbors, the power of the input beam required for soliton formation is smaller than in other configurations, even though the displacement of the integral center is maximal. This becomes apparent when the light evolution is compared to that in an array with a straight planar interface ($\alpha = \pi$), which is shown in Fig. 4. In the linear case, the light again spreads deeply in to the array region. However, the displacement of the integral beam center is substantially smaller than in the wedge-like geometry ($\alpha = \pi/3$) [Fig. 4(a)]. For an increasing input peak power the beam center at the output again moves toward the excited waveguide [Fig. 4(b)]. If the launched beam power exceeds a certain threshold, a discrete surface soliton forms [Fig. 4(c)]. However, the power threshold for $\alpha = \pi$ is considerably higher than in the $\alpha = \pi/3$ case due to increased number of neighboring waveguides and the resulting stronger discrete diffraction. This is in agreement with our simulations.

A sequence of output intensity distributions for increasing input peak powers in a concave geometry ($\alpha = 4\pi/3$) is depicted in Fig. 5. In the low power limit the light spreads into the array region, while the integral beam center is only slightly shifted from the excited waveguide [Fig. 5(a)]. With increasing input peak power, the diffraction spreading is suppressed [Fig. 5(b) and 5(c)]. For sufficiently high input peak powers one can clearly observe near-corner localization [Figs. 5(d) and 5(e)], and finally a corner soliton forms [Fig. 5(f)]. The required threshold power is again larger than in the previous cases, since now there are five neighbors of the excited guide yielding a further enhanced diffraction.

In conclusion, we demonstrated experimentally 2D solitons at the edges joining planar interfaces of a hexagonal waveguide array with a uniform nonlinear medium. In the linear limit the displacement of the integral beam center is maximal for small face angles $\alpha$ and it decreases with growing $\alpha$. In contrast, the threshold power for soliton formation substantially increases with increase of $\alpha$.

# References with titles

# References without titles

# Figure captions

Figure 1.     Microscope image of a laser written array. The excited waveguides are marked with a circle.

Figure 2.     Profiles of surface solitons at (a) $b=0.463$, $\alpha=\pi/3$, (b) $b=0.637$, $\alpha=\pi/3$, and (c) $b=0.476$, $\alpha=4\pi/3$, corresponding to the points marked with circles in the $U(b)$ diagrams (d). The white dashed lines indicate the interface position. In all cases is $p=2.8$.

Figure 3.     Comparison of the output intensity distributions for an excitation of a corner waveguide in an array with $\alpha=\pi/3$. Top row - experiment, bottom row - theory. The input power is (a) $0.15\,\text{MW}$, (b) $1.3\,\text{MW}$, and (c) $3.2\,\text{MW}$.

Figure 4.     Comparison of the output intensity distributions for an excitation of a corner waveguide in an array with $\alpha=\pi$. Top row - experiment, bottom row - theory. The input power is (a) $0.15\,\text{MW}$, (b) $2.5\,\text{MW}$, and (c) $3.5\,\text{MW}$.

Figure 5.     Dynamic excitation of a corner waveguide in an array with $\alpha=4\pi/3$. The input power is (a) $0.15\,\text{MW}$, (b) $1\,\text{MW}$, (c) $2\,\text{MW}$, (d) $2.3\,\text{MW}$, (e) $2.7\,\text{MW}$, and (f) $3.7\,\text{MW}$.

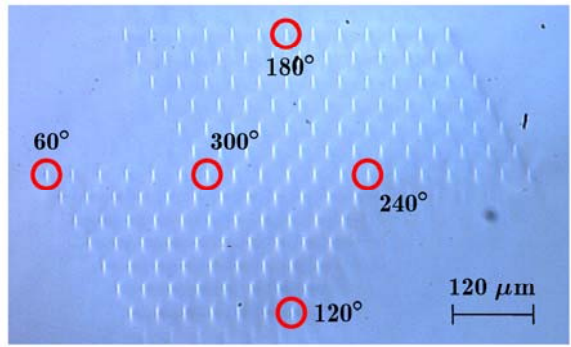

Figure 1. Microscope image of a laser written array. The excited waveguides are marked with a circle.

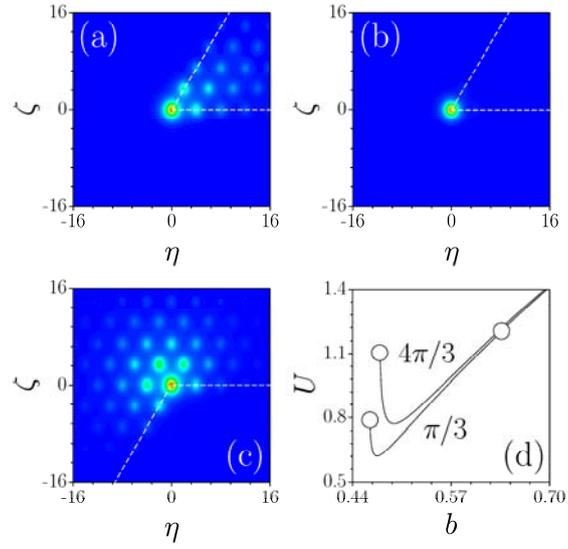

Figure 2. Profiles of surface solitons at (a) $b = 0.463$, $\alpha = \pi/3$, (b) $b = 0.637$, $\alpha = \pi/3$, and (c) $b = 0.476$, $\alpha = 4\pi/3$, corresponding to points marked with circles in the $U(b)$ diagrams (d). White dashed lines indicate interface position. In all cases $p = 2.8$.

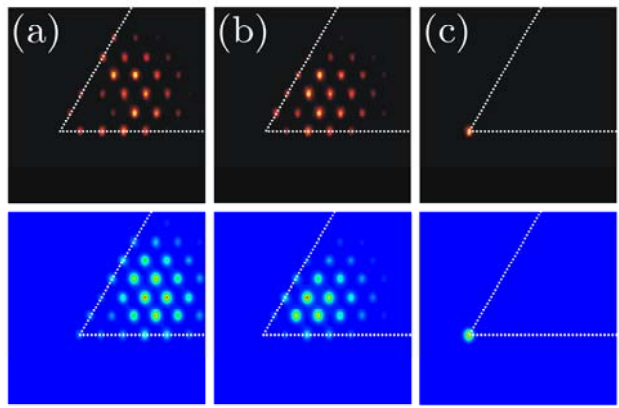

Figure 3. Comparison of the output intensity distributions for an excitation of a corner waveguide in an array with $\alpha = \pi/3$. Top row - experiment, bottom row - theory. The input power is (a) $0.15\,\mathrm{MW}$, (b) $1.3\,\mathrm{MW}$, and (c) $3.2\,\mathrm{MW}$.

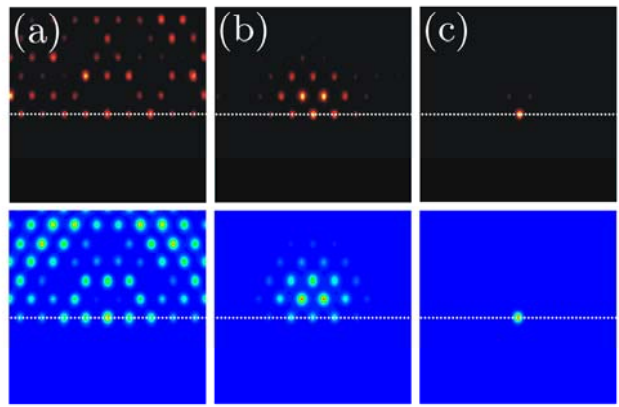

Figure 4. Comparison of the output intensity distributions for an excitation of a corner waveguide in an array with $\alpha = \pi$. Top row - experiment, bottom row - theory. The input power is (a) 0.15 MW, (b) 2.5 MW, and (c) 3.5 MW.

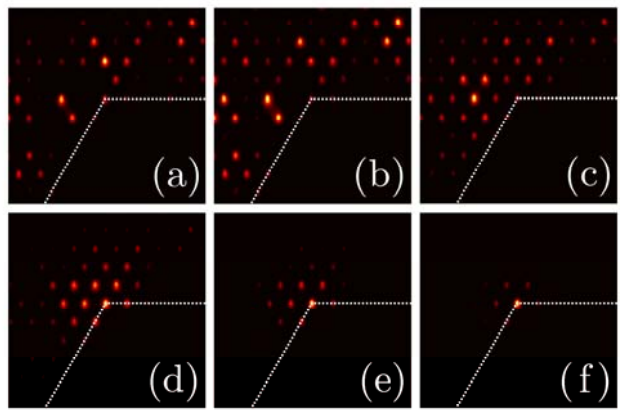

Figure 5. Dynamic excitation of a corner waveguide in an array with $\alpha = 4\pi/3$. The input power is (a) 0.15 MW, (b) 1 MW, (c) 2 MW, (d) 2.3 MW, (e) 2.7 MW, and (f) 3.7 MW.